\documentclass[aps,twocolumn,prl,letterpaper]{revtex4}
\usepackage[pdftex]{graphicx}
\usepackage{amssymb} 
\usepackage{url}     
\usepackage{bm}      

\begin{document}
\newcommand{\ohm}{\ensuremath{\,\Omega}}
\newcommand{\Ef}{\ensuremath{E_\mathrm{F}}}
\newcommand{\Bf}{\ensuremath{B_\mathrm{f}}}
\newcommand{\kf}{\ensuremath{k_\mathrm{F}}}
\newcommand{\rc}{\ensuremath{r_\mathrm{c}}}
\newcommand{\Ii}{\ensuremath{I_\mathrm{i}}}
\newcommand{\Vc}{\ensuremath{V_\mathrm{c}}}
\newcommand{\Lslg}{\ensuremath{L_\mathrm{MLG}}}
\newcommand{\Lblg}{\ensuremath{L_\mathrm{BLG}}}
\newcommand{\Ltlg}{\ensuremath{L_\mathrm{TLG}}}

\setlength{\pdfpageheight}{\paperheight}
\setlength{\pdfpagewidth}{\paperwidth}

\title{Electrically tunable transverse magnetic focusing in graphene}
\author{Thiti  Taychatanapat$^{1,2}$}
\author{Kenji Watanabe$^{3}$}
\author{Takashi Taniguchi$^{3}$}
\author{Pablo Jarillo-Herrero$^{2}$}
\affiliation{$^{1}$Department of Physics, Harvard University, Cambridge, MA 02138, USA}
\affiliation{$^{2}$Department of Physics, Massachusetts Institute of Technology, Cambridge, MA 02139, USA}
\affiliation{$^{3}$National Institute for Materials Science, Namiki 1-1, Tsukuba, Ibaraki 305-0044, Japan}
\date{\today}

\maketitle

{\bf Electrons in a periodic lattice can propagate without scattering for macroscopic distances despite the presence of the non-uniform Coulomb potential due to the nuclei~\cite{Bloch_Blochwave}. Such ballistic motion of electrons allows the use of a transverse magnetic field to focus electrons~\cite{Tsoi_firstMEF}. This phenomenon, known as transverse magnetic focusing (TMF), has been used to study the Fermi surface of metals~\cite{Tsoi_1999} and semiconductor heterostructures~\cite{Houten_ElectronFocusing}, as well as to investigate Andreev reflection~\cite{Tsoi_1999}, spin-orbit interaction~\cite{Rokhinson_SpinSeparation}, and to detect composite fermions~\cite{Goldman_DetectionOfCompositeFermions,Smet_CompositeFermions}. Here we report on the experimental observation of transverse magnetic focusing in high mobility mono-, bi-, and tri-layer graphene devices. The ability to tune the graphene carrier density enables us for the first time to investigate TMF continuously from the hole to the electron regime and analyze the resulting ``focusing fan". Moreover, by applying a transverse electric field to tri-layer graphene, we use TMF as a ballistic electron spectroscopy method to investigate controlled changes in the electronic structure of a material. Finally, we demonstrate that TMF survives in graphene up to $300$ K, by far the highest temperature reported for any system, opening the door to novel room temperature applications based on electron-optics.}

\begin{figure*}
\begin{center}
\includegraphics{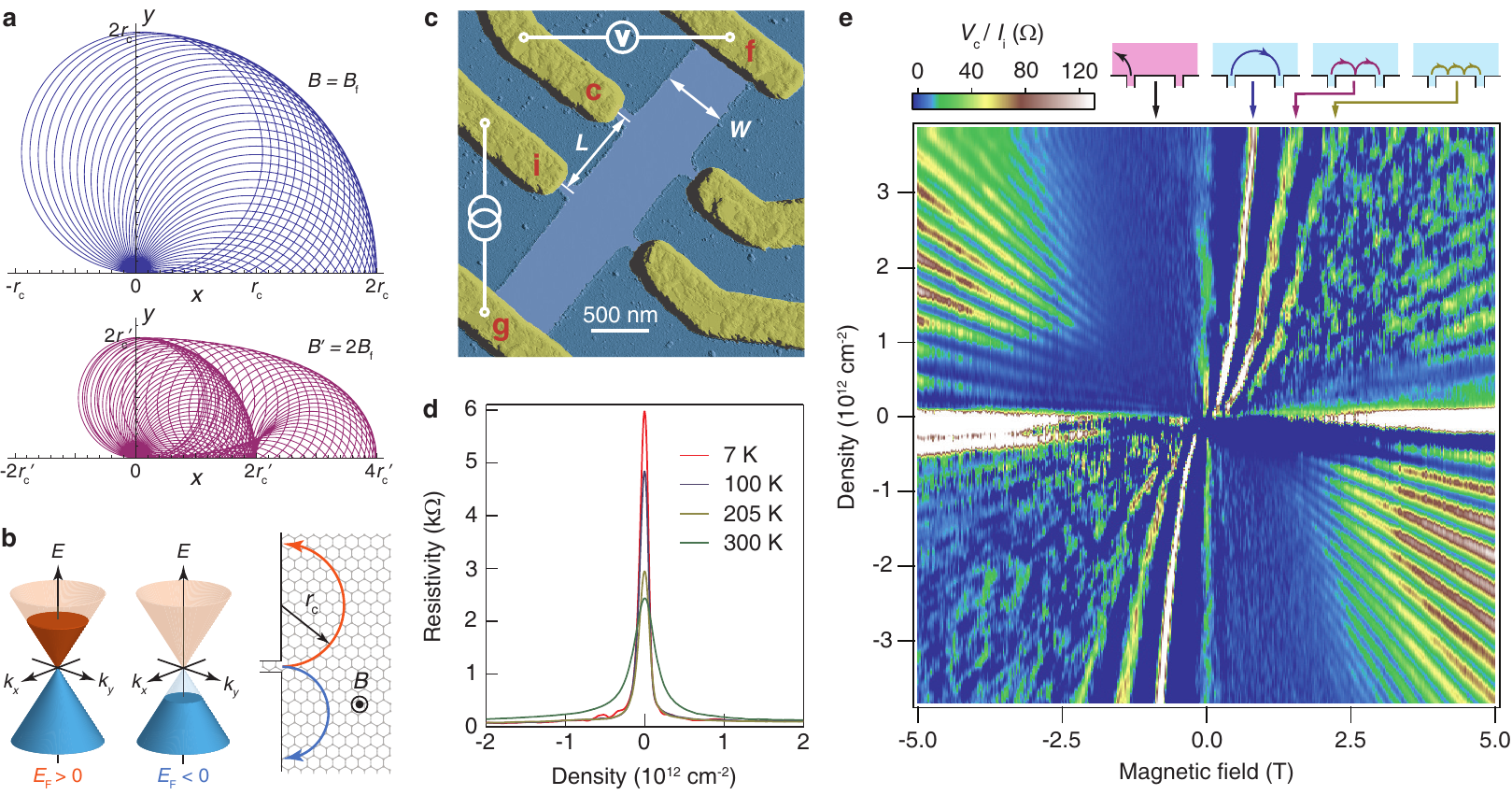}
\caption{{\bf Transverse magnetic focusing schematics.} {\bf a}, Classical trajectories of electrons injected isotropically from the origin at $B = \Bf$ (top) and $B' = 2\Bf$ (bottom, including one bounce off the edge). Electrons are focused at an integer multiple of $2\rc$ along the $x$-axis. {\bf b}, Cartoons depicting the band structure of MLG at positive (left) and negative (center) electron density. Electron's trajectories at a finite $B$ are shown on the right for Fermi energy $\Ef > 0$ (orange line) and $\Ef < 0$ (blue line). {\bf c}, False color atomic force microscopy (AFM) image of a TMF device. In the TMF measurement, contact i injects current $I_i$ into graphene and the voltage $V_c$ is measured at the collector (contact c) relative to contact f. $L$ is the measured distance between the centers of contacts i and c. {\bf d}, Resistance versus gate voltage of MLG at various temperatures measured in the usual 4-probe Hall bar geometry. {\bf e}, TMF spectrum in MLG at $5$ K. TMF peaks from first, second, and third modes can be observed clearly for $|B| < 2.5$ T. The top insets show representative trajectories for each corresponding mode. At higher $B$, SdHOs are also present.} \label{F:Fig1}
\end{center}
\end{figure*}

The concept of TMF can be illustrated by considering electrons entering a two-dimensional system through a narrow injector (origin in Fig. 1a). In the presence of a magnetic field $B$, electrons will undergo cyclotron motion with radius $\rc$ and get focused on the caustic (a quarter of a circle with radius $2\rc$) on which the electron density becomes singular (Fig.~\ref{F:Fig1}a, top).   Moreover, the specular reflection of electrons at the boundary of the two-dimensional system causes a skipping orbit motion which results in focal points at integer multiples of $2\rc$ along the $x$-axis (Fig.~\ref{F:Fig1}a, bottom). This basic behavior still holds for electron motion in a solid as long as the Fermi surface has cylindrical symmetry~\cite{Tsoi_1999}. Hence, the magnetic field, $B_{f}$, required to focus electrons at a distance $L$ is
\begin{equation}
    \Bf^{(p)} = \left(\frac{2 \hbar \kf}{e L} \right)p = \left(\frac{2 \hbar \sqrt{ \pi n}}{e L} \right)p
\end{equation}
where $p-1$ is the number of reflections off the edge of the system (e.g. $p=1$ corresponds to direct injector to collector trajectory, without reflections), $\hbar$ is the reduced Planck's constant, $e$ is the elementary charge, $\kf$ is the Fermi momentum, and where we have used $\kf = \sqrt{\pi n}$, $n$ being the carrier density.

In order to study TMF in graphene, we fabricate Hall bar devices based on high mobility mono- (MLG), bi- (BLG), and tri-layer (TLG) graphene on hexagonal boron nitride (hBN) substrates~\cite{Dean_BN} (see Methods and Fig.~\ref{F:Fig1}c). The multi-terminal geometry required to study TMF imposes a minimum mean free path of the order of several hundred nm, which has only been possible with the advent of G on hBN devices~\cite{Dean_BN,Thiti_tri,Mayorov_BendResistance}. Figure~\ref{F:Fig1}d shows the resistivity of a MLG device as a function of density at zero magnetic field. The device exhibits a narrow Dirac peak with a strong temperature dependence, which indicates low disorder~\cite{Du_Ballistic,Bolotin_TDependentSusGra}. Its field effect mobility is $\sim$$100,000$~cm$^2\,$V$^{-1}\,$s$^{-1}$ at low temperature, corresponding to a mean free path of $\sim$$1$ $\mu$m. A similar behavior is observed for BLG and TLG devices. The high mobility and low disorder enable us to probe TMF in these devices.

We employ the measurement configuration shown in Fig.~\ref{F:Fig1}c to probe the focusing of electrons. Current $\Ii$ is injected through contact i while contact g is grounded and voltage $\Vc$ is measured at the collector (contact c) relative to contact f. The magnetic field is applied normal to graphene. Figure~\ref{F:Fig1}e shows the normalized $\Vc(B,n)$ in MLG, at $5$~K. Two sets of features are immediately apparent: for $|B| \geq 2.5$~T, we observe Shubnikov-de Hass oscillations (SdHOs), forming a usual Landau fan, as expected from the measurement setup, which is topologically equivalent to a longitudinal resistance measurement. While the SdHOs are very pronounced in quadrants 2 and 4 (top-left and bottom-right), they are nearly invisible in quadrants 1 and 3, due to the interference of different trajectories of electrons propagating coherently to the collector~\cite{Houten_ElectronFocusing,Beenakker_modeInterference,Aidala_ImagingFocusing,Rakyta_ElectronFocusingTheory}(see supplementary information).

In the low field regime, $|B| \leq 2.5$~T, we observe three unusual peaks which do not resemble SdHOs. For positive density, these peaks appear on the positive $B$ side. The location of these features in the $B-n$ plane (see Eq.~1) indicates that these peaks can be associated with TMF. The peaks arise when electrons are focused onto the collector, resulting in a build up of $\Vc$. The first peak corresponds to electrons propagating directly from the injector to the collector while for the higher order peaks electrons reflect off the edge before reaching the collector (Fig.~\ref{F:Fig1}e, top insets). For negative $B$, electrons propagate away from the collector and hence no focusing peak is observed (Fig.~\ref{F:Fig1}e, top left inset). As we tune to negative density ($\Ef < 0 $), the sign of the charge carriers flips, and therefore $B$ has to be reversed in order for the carriers to be focused at the collector. The ability to tune density in graphene enables us to investigate the $\sqrt{n}$ dependence of the focusing fields, or ``focusing fan'', continuously from the electron to hole regimes in a single device over a broad density range, which was never done in other systems.

\begin{figure*}
\begin{center}
\includegraphics{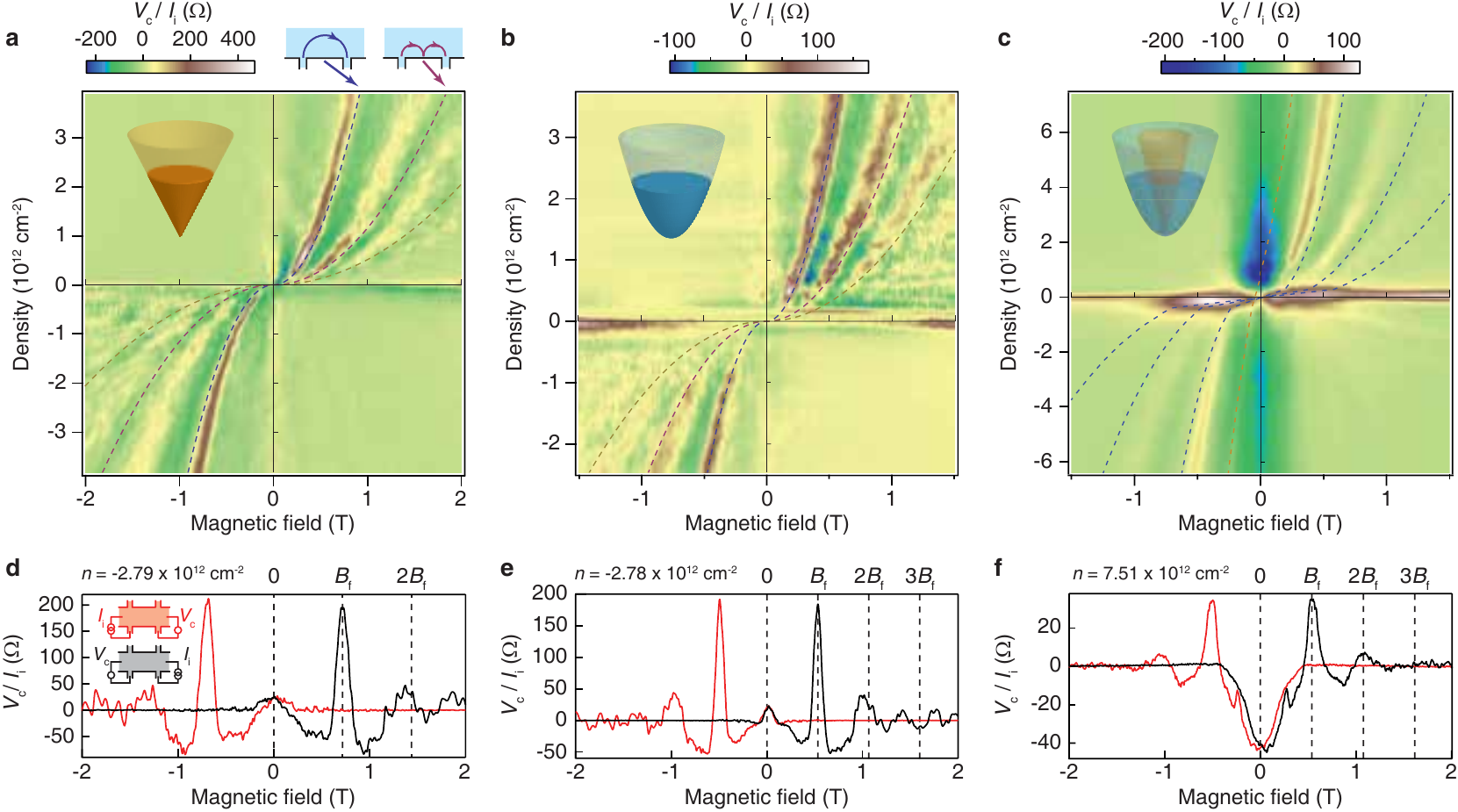}
\caption{{\bf transverse magnetic focusing in MLG, BLG, and TLG at $\mathbf{5}$ K.} {\bf a-c}, The TMF spectra as a function of density and magnetic field for MLG, BLG, and TLG respectively. The dashed lines are calculated focusing fields using $\Lslg = 500$, $\Lblg = 775$, and $\Ltlg = 950$~nm, determined from AFM images. {\bf d-f},  Onsager reciprocal relation in MLG, BLG, and TLG respectively. The red and black traces are measured with current and voltage contacts switched (Fig.~\ref{F:Fig2}d, insets). In these figures, $\Bf$ is the observed focusing field of the first mode.} \label{F:Fig2}
\end{center}
\end{figure*}

The values of $\Bf$ can be readily calculated, since both $n$ and $L$ can be obtained from Hall measurements and the AFM image of the device, respectively. Figure~\ref{F:Fig2}a shows a zoom-in plot of Fig.~\ref{F:Fig1}e, where we have superimposed the calculated focusing fields (dashed lines) using the measured $\Lslg = 500$~nm. A discrepancy between the calculated values and the measured peak locations is clearly present. Moreover, we find that the observed $\Bf^{(p)}/p$ decreases as $p$ increases (Fig.~\ref{F:Fig2}d). The finite width of our injector and collector ($\sim$$100$~nm in MLG and BLG and $\sim$$240$~nm in TLG, see below) could introduce an error in the determination of $L$ and subsequently $\Bf$. However, a more plausible explanation is the effect of charge accumulation near the edges owing to the finite size of our graphene devices~\cite{Efetov_ChargeAccumulation}. We find that charge accumulation reduces $\Bf$ by the same order of magnitude as that required to correct for the discrepancy and, in addition, it also explains the decreasing $\Bf^{(p)}/p$ because, for higher $p$, the carrier's trajectory is closer to the edge, which further reduces $\Bf^{(p)}$ (see supplementary information). We also note that density fluctuations and small-angle scattering~\cite{Aidala_ImagingFocusing} due to impurities could also affect the carrier's path and its focus. However, a lack of knowledge of the detailed disorder potential landscape prevents us from determining the change in the value of the focusing fields.

We have observed multiple focusing peaks in all of our devices, including BLG and TLG (see below), which indicates that a significant fraction of the electrons get specularly reflected off the graphene edge. From the peak amplitudes, we can calculate the measured specularity, the ratio between the amplitude of the second mode to that of the first mode, which offers information on the specular reflection of electrons at the graphene edges. We find that the value of  specularity ranges from 0.2 to 0.5 (see supplementary information). It is worth noting that specularity measurements in semiconductor heterostructures have shown values less than 1 for focused-ion-beam etched devices which is similar to our oxygen-plasma-etched graphene devices but greater than 1 for electrostatically-defined edges~\cite{Tsoi_1999,Houten_ElectronFocusing}.

\begin{figure*}
\begin{center}
\includegraphics{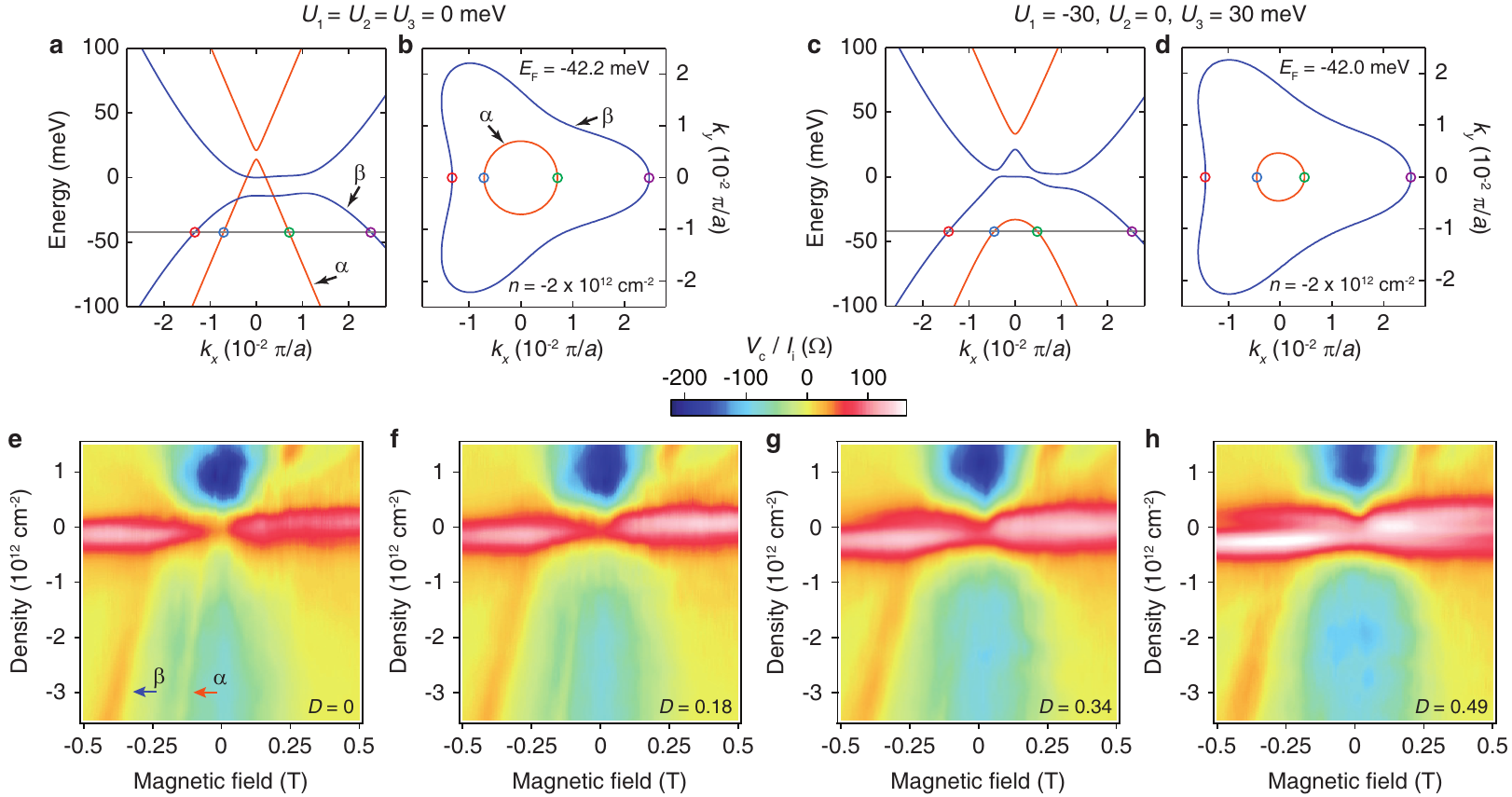}
\caption{{\bf Visualizing in-situ band structure changes in TLG with TMF.} {\bf a-b}, Band structure and Fermi surface of TLG at zero $D$ (electrostatic potential of each layer equal to zero). The band structure consists of MLG-like and BLG-like subbands, with a small band overlap. The bands $\alpha$ and $\beta$ are MLG-like and BLG-like valence bands. The trigonal warping effect can be seen in the BLG-like subband. The lattice constant $a$ is $2.46$ {\AA}.  {\bf c-d}, Band structure and Fermi surface of TLG at finite $D$ (for this case, with potential difference between adjacent layers equal to $30$~meV). The potential difference induces the hybridization between MLG-like and BLG-like subbands and also shifts down in energy the top of the $\alpha$ band. {\bf e-h}, The TMF spectra in TLG at $D = 0$, $0.18$, $0.34$, and $0.49$~V/nm respectively. As $D$ increases, the $\alpha$ band starts to disappear while the $\beta$ band remains visibly unchanged.} \label{F:Fig3}
\end{center}
\end{figure*}

We now turn to TMF in BLG. Figures~\ref{F:Fig2}b and e display TMF fans for BLG at $5$~K. Evidently, the TMF spectra of MLG and BLG are very similar, even though their band structures are different (Fig.~\ref{F:Fig2}a-b, insets). The similarity arises from the fact that, when only the nearest intra-layer $\gamma_0$ and inter-layer $\gamma_1$ hopping parameters are considered, both MLG and BLG have circular Fermi surfaces, resulting in the same circular orbit and $\sqrt{n}$-dependence of $\kf$. The dashed lines in Fig.~\ref{F:Fig2}b are focusing fields calculated from Eq.~1. For this device, the calculated $\Bf^{(1)}$ for the first order peak are in good agreement with the measured data, but higher order modes show a discrepancy, similar to the MLG case. An additional possible source of mismatch in BLG, which does not exist in MLG, is trigonal warping~\cite{McCann_BLG_g3} of the Fermi surface due to the next nearest neighbor inter-layer hopping term $\gamma_3$. This term transforms the BLG circular Fermi surface into a partly triangular surface, altering therefore the carrier's trajectory. Hence, in principle the values of $\Bf$ now depend on the crystallographic orientation with respect to the sample axis, and can vary by a few tens of mT (see supplementary information).

Even though TMF cannot be used to differentiate MLG from BLG, the TMF spectrum of TLG is remarkably different because of the multiband character of its band structure. Figure~\ref{F:Fig2}c and f show TMF spectra of TLG at $5$~K, measured at zero electric displacement field. Taking only $\gamma_0$ and $\gamma_1$ into account, the band structure of TLG consists of a massless MLG-like and a massive BLG-like subband at low energy (Fig.~\ref{F:Fig2}c, inset)~\cite{Thiti_tri,Lu_ABAbandStructure,Guinea_ABAbandStructure,Latil_ABAbandStructure,Partoens_ABAbandStructure}. In a magnetic field, both subbands give rise to their own TMF spectra, with the BLG-like subband having a larger $\Bf^{(1)}$ due to its larger Fermi momentum (for a given $\Ef$) . This allows us to identify the subband corresponding to each peak observed in the data.  At high density, the peak from the MLG-like subband can be seen at $\sim$$250$~mT (Fig.~\ref{F:Fig2}c, orange dashed line, and small sharp peaks at low field in Fig.~\ref{F:Fig2}f) while the peaks from the BLG-like subband are visible from $\sim$$250$~mT onward (Fig.~\ref{F:Fig2}c, blue dashed lines). We do not observe higher order modes from the MLG-like subband, probably because they are masked by the much stronger peaks from the BLG-like subband, which contains most of the charge density.

Earlier studies have shown that higher order hopping parameters in TLG significantly modify its band structure~\cite{Thiti_tri,Lu_ABAbandStructure,Guinea_ABAbandStructure,Latil_ABAbandStructure,Partoens_ABAbandStructure} by introducing subband overlap and trigonal warping in the BLG-like subband (Fig.~\ref{F:Fig3}a). We use this full parameter model for the TLG band structure to simulate the carrier trajectories and determine the focusing fields (see supplementary information). The results are shown as dashed lines in Fig~\ref{F:Fig2}c. Although we can reproduce the focusing field for the MLG-like subband very accurately, we obtain a mismatch in the BLG-like subband, similar to those above mentioned in MLG and BLG.

We now focus on the previously unexplored potential of TMF as a ballistic electron spectroscopy method to investigate in-situ changes in the band structure of a material. One of the remarkable properties of TLG is that its band structure can be tuned and controlled by using a transverse electric displacement field~\cite{Koshino_ABA_Gate}, $D$. TMF is sensitive to the occupation of each of the TLG subbands, enabling us to use TMF as a probe of the change in the TLG band structure with $D$. Figures~\ref{F:Fig3}a and b show the band structure and Fermi surface of TLG at $n = -2 \times 10^{12}$~cm$^{-2}$ for the case $D=0$. We denote the valence bands of the MLG-like and BLG-like subbands as $\alpha$ and $\beta$ bands, respectively. The application of a finite $D$ induces a potential difference between the TLG layers, breaking the mirror symmetry and causing a hybridization between the MLG-like and BLG-like subbands. Figures~\ref{F:Fig3}c and d show how a finite $D$ results in a shift down of the top of the $\alpha$ band . Consequently, for a fixed density, the Fermi momentum of the $\alpha$ band shrinks with $D$ while that of the $\beta$ band barely changes due to its much higher density of states.

Figures~\ref{F:Fig3}e-h show the TMF spectra of TLG at various $D$'s. We observe a relatively strong focusing peak from the $\alpha$ band at $D=0$~V/nm. However, as $D$ increases, the peak starts to shift downward and it eventually disappears at low density. The disappearance of this peak is the result of the top of the $\alpha$ band shifting down in energy and leaving the $\beta$ band as a lone contributor to the carrier density (Fig.~\ref{F:Fig3}c-d). Therefore, within our density range, we end up observing only a single focusing peak, from the $\beta$ band, at high $D$. The onset in the density of the focusing peak of the $\alpha$ band allows us to determine the potential difference among the TLG layers as a function of applied $D$.  As a result, we can estimate the effective dielectric constant of TLG which we find to be about $3.5\pm0.2$ (see supplementary information).

\begin{figure*}
\begin{center}
\includegraphics{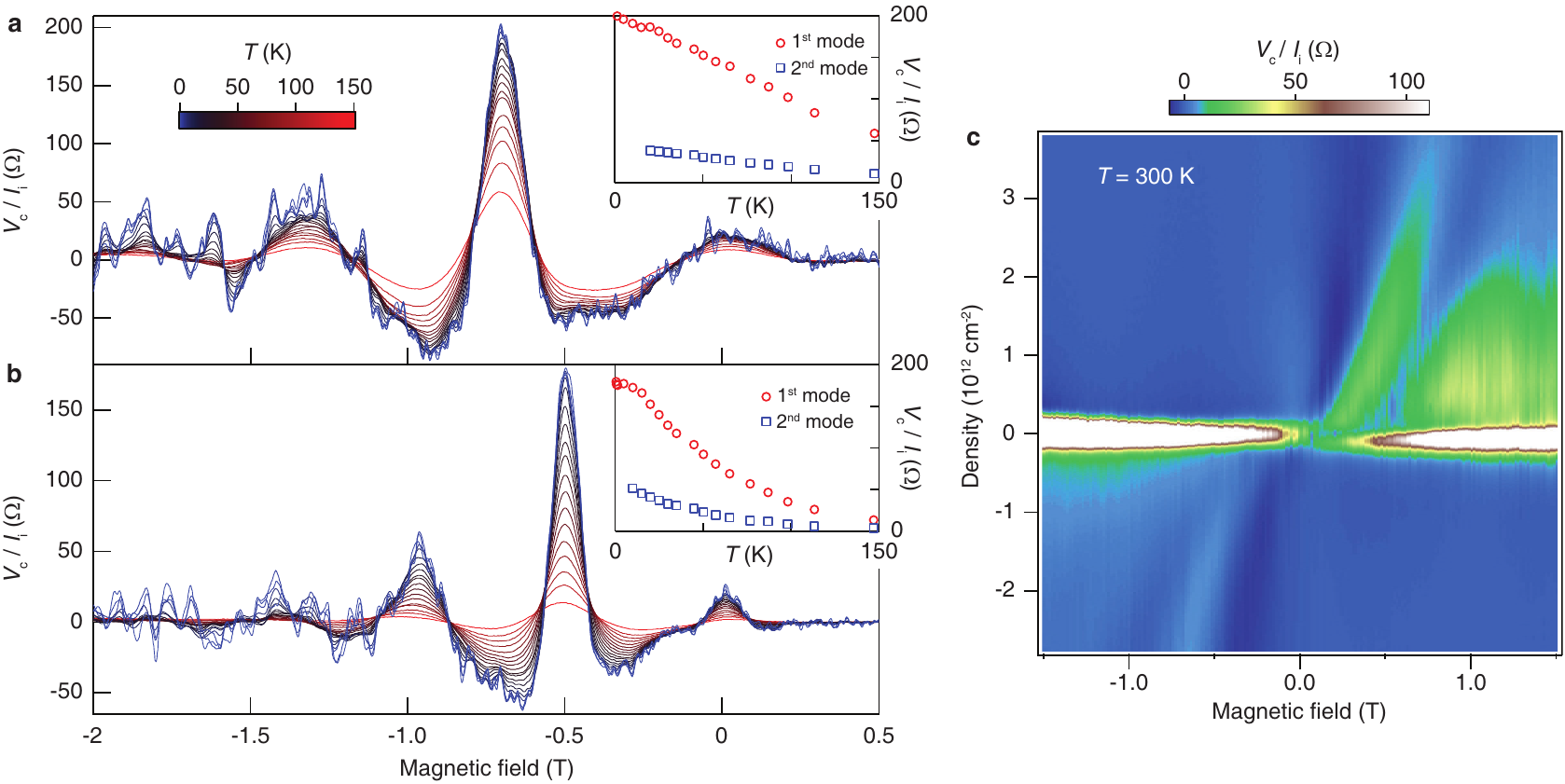}
\caption{{\bf Temperature dependence of the TMF in MLG and BLG.} {\bf a-b} The TMF spectra as a function of temperature from $300$~mK to $150$ K for MLG and BLG, respectively, at $n = -2.8 \times 10^{12}$~cm$^{-2}$. Insets show the amplitudes of the first and second modes as a function of temperature (Data taken before current annealing). {\bf c} transverse magnetic focusing in MLG at $300$~K. The TMF peak as well as the $\sqrt{n}$ dependence of the focusing field can be clearly observed (Data taken after current annealing).} \label{F:Fig4}
\end{center}
\end{figure*}

We now look at the temperature dependence of the TMF spectra in MLG and BLG. The TMF spectrum is affected by temperature, $T$, at least in two ways: through the increase in dephasing (which smoothes the quantum interference fluctuations), and through the loss of ballistic transport due to new scattering channels activated at high $T$. Figures~\ref{F:Fig4}a and b show the TMF spectra of MLG and BLG, respectively, at $n = -2.8 \times 10^{12}$~cm$^{-2}$ from $0.3$ to $150$~K. We first concentrate on the fine structure observed at low $T$ (blue traces). This structure is the aforementioned quantum interference between different paths on which electrons propagate to the collector~\cite{Houten_ElectronFocusing,Beenakker_modeInterference,Aidala_ImagingFocusing,Rakyta_ElectronFocusingTheory}. When the temperature-induced broadening of the Fermi momentum is on the order of $1/L$, electrons become incoherent and the quantum interference is washed out~\cite{Cheianov_VeselagoLens}, resulting in smooth focusing peaks. For our devices, this length corresponds to a temperature of about $15$~K, which is in good agreement with the data.

In addition, the focusing peaks also decrease as $T$ increases. The amplitudes of the first and second modes are shown in the insets of Fig.~\ref{F:Fig4}a and b for MLG and BLG, respectively. We observe that the focusing amplitude in MLG depends linearly on $T$. A potential scattering mechanism includes longitudinal acoustic phonons, which give rise to a linear temperature dependence of the scattering rate~\cite{Hwang_AcousticPhonon}. However, we observe a very different temperature dependence in BLG. The peak amplitude saturates at low temperature and decreases faster than in MLG at higher $T$. A similar temperature dependence of the focusing peaks has also been observed in InGaAs/InP heterojunctions~\cite{Heremans_TDependence}. Further theoretical work is needed to understand the temperature dependence of the focusing peaks as well as the difference between MLG and BLG.

We end by commenting on the remarkable robustness of TMF in graphene. Fig.~\ref{F:Fig4}c shows the TMF fan of MLG at room temperature ($T=300$~K), where the first mode is clearly visible, indicating room temperature ballistic transport well into the micron regime. This lower bound temperature for the observation of TMF in graphene is at least three times higher than the highest temperature at which TMF spectrum has been observed in semiconductor heterostructures~\cite{Heremans_TDependence}, the main reason probably being the lack of remote interfacial phonon scattering~\cite{Guinea_TDepend_hBN} from hBN. The ability to manipulate ballistic motion in graphene at room temperature, coupled with recent developments~\cite{Liu_GrapheneOnhBN} in large area growth of graphene on hBN, paves the way towards novel applications based on electron-optics. On a more fundamental level, TMF may serve as a probe of electron-electron interaction~\cite{Novoselov_blg_interaction, Weitz_bilayer,CastroNeto_eeInteraction} or strain-induced gauge field~\cite{Guinea_strain, Crommie_Strain, Manoharan_graphene} effects in the electronic structure of graphene.

\subsection*{Methods}
Figure 1c shows an atomic force microscopy image from one of our devices. Our devices are fabricated by transferring graphene onto high-quality hexagonal boron nitride~\cite{Dean_BN}. We use oxygen plasma to etch graphene flakes into a Hall-bar geometry. Contacts are defined by electron-beam lithography and thermal evaporation of chromium and gold. The devices are then heat annealed in forming gas and subsequently current annealed in vacuum at low temperature~\cite{Thiti_tri}. We observe TMF peaks both before and after current annealing. The data after current annealing have higher quality than before current annealing, especially at low density, likely due to reduced charge density fluctuations. However, they exhibit similar quality at high density. All the data shown here are measured after current annealing, except Fig.~\ref{F:Fig2}d-e and Fig.~\ref{F:Fig4}a-b which was done before current annealing.

We identify the number of graphene layers by Raman spectroscopy and/or quantum Hall measurements. For TLG, the quantum Hall measurements reveal that it is Bernal-stacked~\cite{Thiti_tri}. In addition, we put a top gate onto the TLG device, using hBN as a thin dielectric. The combination of top gate (TG) and bottom gate (BG) allows us to control the charge density and the displacement field independently. We parameterize the displacement field by $D = (C_{\mathrm{TG}} V_{\mathrm{TG}} + C_{\mathrm{BG}}V_{\mathrm{BG}})/(2 \epsilon_0)$ where $C$ is the capacitive coupling, $V$ is the applied gate voltage relative to the charge neutrality point, and $\epsilon_0$ is the vacuum permittivity.

\subsection*{Acknowledgements}
We thank L. Levitov  and A. Yacoby for discussions. We acknowledge financial support National Science Foundation Career Award No. DMR-0845287 and the Office of Naval Research GATE MURI. This work made use of the MRSEC Shared Experimental Facilities supported by the National Science Foundation under award No. DMR-0819762 and of Harvard's Center for Nanoscale Systems (CNS), supported by the National Science Foundation under grant No. ECS-0335765.

\subsection*{Author contributions}
T. Taychatanapat fabricated the samples and performed the experiments. K.W. and T. Taniguchi synthesized the hBN samples. T. Taychatanapat and P.J-H. carried out the data analysis and co-wrote the paper.

\subsection*{Additional information}
Correspondence and requests for materials
should be addressed to P.J-H.~(email: \mbox{pjarillo@mit.edu})

\end{document}